\documentclass[aps,prd,onecolumn,superscriptaddress,nofootinbib,preprintnumbers,11pt]{revtex4-2}
\usepackage{amsmath,amssymb,graphicx,hyperref,xcolor}
\usepackage{multirow,booktabs}
\usepackage{physics}
\usepackage{siunitx}
\usepackage{slashed}
\usepackage{url}
\usepackage{bm}

\begin{document}

\title{Machine-Learning-Inspired SMEFT Simplified Template Cross Sections:\\
A Case Study in \texorpdfstring{$ZH$}{ZH} Production}

\author{Daniel Conde}
\affiliation{Instituto de Física Corpuscular (IFIC), CSIC--Universitat de Val\`encia, Spain}
\author{Miguel G. Folgado}
\affiliation{Instituto de Física Corpuscular (IFIC), CSIC--Universitat de Val\`encia, Spain}
\author{Veronica Sanz}
\affiliation{Instituto de Física Corpuscular (IFIC), CSIC--Universitat de Val\`encia, Spain}

\date{\today}

\begin{abstract}
The Simplified Template Cross Section (STXS) program has become the standard interface between Higgs measurements and global fits, but its fixed one-dimensional boundaries are not guaranteed to align with the phase-space directions to which the Standard Model Effective Field Theory (SMEFT) is most sensitive. We propose a machine-learning-inspired extension of STXS in which supervised classifiers are used only at the design stage to identify simple, publishable phase-space boundaries. Using associated Higgs production, $pp\to ZH$, as a case study and a benchmark momentum-dependent bosonic SMEFT deformation, we show that the relevant signal–background separation is well captured by a linear boundary in the $(p_T^Z,m_{ZH})$ plane. We construct such boundaries with a linear support vector machine and with a deep-neural-network-assisted distillation procedure, and compare them directly with the standard STXS $p_T^Z$ bins through a common single-region Asimov-significance analysis. In this proof-of-concept setup, the ML-inspired regions systematically outperform the corresponding STXS regions, with the largest gains appearing in the boosted regime where SMEFT effects are concentrated. The final observable remains a simple linear cut, preserving the transparency and experimental portability that make STXS useful.
\end{abstract}

\maketitle
\tableofcontents

\section{Introduction}
\label{sec:intro}

The Higgs sector remains one of the sharpest probes of physics beyond the Standard Model (SM). In the absence of direct evidence for new particles, the Standard Model Effective Field Theory (SMEFT) provides a systematic language in which small departures from the SM can be organized through higher-dimensional operators suppressed by a heavy scale $\Lambda$. Global analyses of Higgs, electroweak, top, and diboson data now routinely use the SMEFT framework to translate precision measurements into constraints on Wilson coefficients; representative examples include Refs.~\cite{Ellis:2020unq,Falkowski:2019hvp}. Complementary fitting frameworks and EFT-translation tools commonly used in this context include {\tt fitmaker}, {\tt SMEFiT}, {\tt HEPfit} and {\tt SFitter}~\cite{Ellis:2020unq,Giani:2023gfq,HEPfit,SFitter}.

Among Higgs production modes, associated production with an electroweak vector boson is especially attractive for SMEFT studies because several bosonic operators induce derivative $hVV$ interactions whose effects grow with energy. This feature was already emphasized in early studies of boosted $VH$ production~\cite{Ellis:2012xd,Ellis:2013ywa,Ellis:2014dva,Ellis:2014jta} and was later quantified in SMEFT analyses of momentum-dependent Higgs--gauge interactions~\cite{Azatov:2012bz,Azatov:2013hya,Bishara:2017etm,Biekotter:2018jzu,Biekotter:2019kde,Brivio:2019myy}. In practice, the relevant distortions populate a correlated high-energy region rather than a single kinematic tail: events migrate simultaneously toward large vector-boson transverse momentum and large invariant mass of the $VH$ system.

The experimental language used to report Higgs measurements, however, is the Simplified Template Cross Section (STXS) framework~\cite{deFlorian:2016spz,ATLAS:STXS,CMS:STXS,Hays:2673969}. STXS is successful precisely because it is transparent, experimentally robust, and easy to reinterpret. But that robustness comes with a structural limitation: the regions are predefined and typically built from one-dimensional slices such as $p_T^V$. In a channel like $ZH$, where the SMEFT deformation is intrinsically correlated in $(p_T^Z,m_{ZH})$, such bins need not be aligned with the most informative direction in phase space.

Machine learning offers an obvious route to improved discrimination. Supervised classifiers can exploit multidimensional correlations and approximate likelihood-ratio information far beyond what is available in simple cut-based analyses~\cite{Brehmer:ML,Brehmer:2018kdj,Cranmer:2015bka,Kasieczka:NNinterpret}. Within this approach, however, we aim for portability over power. A nonlinear classifier output is not, by itself, a good public observable: output like this depends on training details, architecture choices, and internal calibrations, and is therefore poorly matched to the role played by STXS in the experimental-theory interface. The $VH$ study of Ref.~\cite{Freitas:2019hbk} illustrates this tension well, where we see that the classifier is powerful, but its output is not a measurement definition one would naturally publish as an STXS stage.

The central idea of this paper is to use machine learning only as a \emph{design tool}. Rather than publishing an opaque discriminator, we let the classifier identify the physically relevant phase-space direction and then distill that information into a simple linear boundary in a low-dimensional plane. In the present case, the target plane is $(p_T^Z,m_{ZH})$, chosen because it captures the dominant high-energy deformation induced by momentum-dependent Higgs--gauge operators and because both variables are experimentally natural.

We develop this idea as a proof of concept in $pp\to ZH$, using a benchmark in which a single bosonic Wilson coefficient is switched on. We consider two constructions. The first is a purely linear support vector machine (SVM) trained directly on $(p_T^Z,m_{ZH})$. The second uses a deep neural network (DNN) trained on a broader feature set and then distills its high-score region into a linear boundary in the same two-dimensional plane. In both cases the final object is not the classifier score but a straight line, directly comparable to the official STXS boundaries.

Our statistical comparison is deliberately narrow and transparent. We compare each standard STXS region to an ML-inspired region within the same $p_T^Z$ slice, and evaluate the resulting single-region sensitivity with the Asimov significance, including representative fractional background uncertainties. This paper therefore does \emph{not} propose a full replacement for the STXS program, nor does it present a combined experimental analysis. It establishes a more modest point: even when one insists on a simple publishable region definition, ML-guided boundaries can recover a meaningful fraction of the multidimensional information lost by fixed one-dimensional STXS cuts.

The rest of the paper is organized as follows. In Sec.~\ref{sec:context} we summarize the SMEFT motivation and the STXS baseline relevant for $ZH$ production. Section~\ref{sec:setup} describes the event generation, feature construction, and statistical setup. In Sec.~\ref{sec:ml} we introduce the ML-inspired region-construction strategies based on a linear SVM and on DNN distillation. Section~\ref{sec:results} presents the boundary comparisons and the significance results. We discuss the scope and limitations of the method in Sec.~\ref{sec:conclusions}. Technical material is collected in the appendices.

\section{Physics target and STXS baseline}
\label{sec:context}

\subsection{SMEFT origin of the boosted deformation}
\label{subsec:eft}

At dimension six, the SMEFT Lagrangian can be written as
\begin{equation}
\mathcal{L}_{\rm SMEFT}=\mathcal{L}_{\rm SM}+\sum_i\frac{c_i}{\Lambda^2}O_i.
\end{equation}
For associated Higgs production, a particularly relevant class of deformations is generated by bosonic operators that modify the momentum structure of the $hVV$ vertex. In the Warsaw basis these include
\begin{align}
O_{HW}  &= (H^\dagger H)\,W^i_{\mu\nu}W^{i\mu\nu}, \\
O_{HB}  &= (H^\dagger H)\,B_{\mu\nu}B^{\mu\nu}, \\
O_{HWB} &= (H^\dagger\tau^i H)\,W^i_{\mu\nu}B^{\mu\nu}.
\end{align}
After electroweak symmetry breaking these operators induce derivative interactions of schematic form $hV_{\mu\nu}V^{\mu\nu}$. Their contribution to the amplitude therefore grows with the characteristic event energy, which drives SMEFT effects into the boosted tails of associated production~\cite{Ellis:2012xd,Ellis:2013ywa,Ellis:2014dva,Ellis:2014jta,Azatov:2012bz,Azatov:2013hya,Biekotter:2018jzu,Bishara:2017etm,Biekotter:2019kde,Brivio:2019myy}.

Appendix~\ref{app:hZZcoupling} gives the corresponding effective $hZZ$ vertex. The three operators above generate the same Lorentz structure up to electroweak mixing factors. This does \emph{not} prove that every optimized boundary is identical across operator space, but it does explain why the dominant deformation should generically track a correlated high-energy direction. In this paper we work with a benchmark in which only $c_{HW}/\Lambda^2$ is nonzero; our claims are therefore restricted to a proof-of-concept demonstration in that setting.

These considerations motivate the use of $p_T^Z$ and $m_{ZH}$ as the primary variables for region design. The first tracks the boosted electroweak recoil and the second captures the hard scale of the full associated system. As we will see, most of the useful phase-space information can be organized in this two-dimensional plane.

\subsection{Why STXS is the right baseline}
\label{subsec:stxs}

The STXS framework was introduced to provide theory-friendly and experimentally stable Higgs measurements~\cite{deFlorian:2016spz} and is now standard in ATLAS and CMS analyses~\cite{ATLAS:STXS,CMS:STXS}. Instead of reporting only inclusive rates, STXS partitions the production phase space into exclusive bins with limited extrapolation dependence. For $ZH$ production the relevant Stage-1/1.2 categories are organized in intervals of the vector-boson transverse momentum,
\begin{eqnarray}
75~\text{GeV} &<& p_T^Z < 150~\text{GeV}, \qquad
150~\text{GeV} < p_T^Z < 250~\text{GeV}, \nonumber \\
250~\text{GeV} &<& p_T^Z < 400~\text{GeV}, \qquad
p_T^Z > 400~\text{GeV}.
\label{eq:stxsbins}
\end{eqnarray}
These bins are routinely used in SMEFT reinterpretations and therefore form the natural baseline for the present study.

We see that in STXS the $p_T^Z$ slicing is structurally one-dimensional, and we suspect that in a channel where the relevant deformation occupies a correlated region of large $p_T^Z$ and large $m_{ZH}$, a vertical cut in $p_T^Z$ cannot be expected to be optimal. The aim of this work is therefore not to bypass STXS but to ask whether one can preserve its experimental virtues while allowing the region boundaries themselves to be informed by the data morphology.

\section{Simulation, observables, and statistical setup}
\label{sec:setup}

\subsection{Event generation and benchmark definition}
\label{subsec:generation}

We consider
\[
pp\to ZH \to \ell^+\ell^- b\bar b,
\]
with $\ell=e,\mu$ at $\sqrt{s}=13~\text{TeV}$. Events are generated with \textsc{MadGraph5\_aMC@NLO}~\cite{Alwall:2014hca} using the \textsc{SMEFTsim} implementation of the Warsaw-basis dimension-six operators~\cite{Brivio:2017bnu}. We use NNPDF3.1~\cite{Ball:2017nwa} throughout. 

To keep the proof-of-concept setup as controlled as possible, we compare an SM sample to a benchmark SMEFT sample with
\begin{equation}
\frac{c_{HW}}{\Lambda^2}=1~\text{TeV}^{-2},
\end{equation}
and all other Wilson coefficients set to zero.

In the statistical treatment below we use the SM sample as the reference expectation $b$ and the EFT deformation relative to it as the excess $s$. The significance quoted in this paper should therefore be read as a shape-sensitive \emph{benchmark discrimination metric} inside a given region. This in contrast to a fully realistic discovery significance against the complete set of experimental backgrounds. This distinction is essential to the interpretation of the results.

\subsection{Event selection and observable set}
\label{subsec:selection}

Events are required to contain two isolated, opposite-sign, same-flavor leptons reconstructing the $Z$ boson and two $b$-tagged jets compatible with the Higgs decay. Standard acceptance and reconstruction cuts are applied to leptons and jets.

From each event we construct a feature set containing the core kinematic information of the process:
\begin{itemize}
    \item transverse momenta: $p_T^Z$, $p_T^H$, $p_T^{\ell_1}$, $p_T^{\ell_2}$, $p_T^{b_1}$, $p_T^{b_2}$, $p_T^{ZH}$;
    \item transverse mass: $m_T^{ZH}$;
    \item invariant masses: $m_{\ell\ell}$, $m_{bb}$, $m_{\mathrm{inv}}^{\mathrm{tot}}$;
    \item angular variables: $\eta_{H}$, $\Delta R_{\ell\ell}$, $\Delta R_{jH}$, and selected $\Delta\phi_{\ell_i b_j}$ combinations;
\end{itemize}

Many of these observables are correlated. The purpose of the DNN is precisely to test whether those extra correlations materially change the final region design after projection into $(p_T^Z,m_{ZH})$. A principal-component analysis is presented in Appendix~\ref{app:PCA}; it is consistent with the empirical observation that the dominant SMEFT--SM variation is strongly tied to an energy-like direction in feature space.

\subsection{Training samples and feature preprocessing}
\label{subsec:dataset}

Each event is represented by a feature vector $x$ and labeled as SM ($y=0$) or SMEFT benchmark ($y=1$). The full dataset contains $2\times 10^6$ events. To estimate the stability of the pipeline, we repeat the full analysis over an ensemble of independent random train/test partitions generated from different random seeds. For each repetition, the data are split into training, validation, and test subsets in the ratio $70\%:10\%:20\%$. Whenever results in this paper are shown with distributions or quoted uncertainties, these correspond to the spread across that ensemble of repeated splits. The results shown in Section~\ref{sec:results} are calculations of each model performed on respective unseen data from the corresponding test set.

Input features are standardized to zero mean and unit variance. Event weights are kept throughout so that the classifiers are trained with the correct physical class weighting rather than on an artificially balanced sample. The linear SVM is trained only on $(p_T^Z,m_{ZH})$. The DNN is trained on the broader feature set of Sec.~\ref{subsec:selection}, but its output is never proposed as the final public observable, but instead projected down to the $(p_T^Z,m_{ZH})$ from its output via SVM.

\subsection{Single-region statistical metric}
\label{subsec:asimov}

All comparisons in this paper are performed \emph{region by region}. For any one region---an official STXS bin, an SVM-defined region, or a DNN-distilled region---we evaluate the sensitivity with the Asimov significance~\cite{Cowan:Asimov}. Without systematics,
\begin{equation}
Z_A=\sqrt{2\left[(s+b)\ln\left(1+\frac{s}{b}\right)-s\right]}.
\label{eq:asimov_basic}
\end{equation}
Including a fractional background uncertainty $\epsilon=\sigma_b/b$, we use the profiled approximation
\begin{equation}
Z_A^2 = 2 \left[(s+b)\ln\!\frac{(s+b)(b+\sigma_b^2)}{b^2+(s+b)\sigma_b^2}
-
\frac{b^2}{\sigma_b^2}\ln\!\left(1+\frac{s\,\sigma_b^2}{b(b+\sigma_b^2)}\right)\right],
\label{eq:asimov_syst}
\end{equation}
with $\sigma_b=\epsilon b$. We study the representative values
\begin{equation}
\epsilon\in\{0.3,0.5,1.0\}.
\end{equation}

The benchmark values are not chosen arbitrarily: they are intended to span the order of magnitude of current uncertainties encountered across  VH measurements and STXS-like regions, especially once one moves to statistically limited or highly boosted corners of phase space. In this sense, $\epsilon$ should be interpreted as an effective analysis-level nuisance scale, not as a detailed decomposition of experimental systematics.

This choice is intentionally simple. We do not attempt a combined multi-bin likelihood or a full nuisance-parameter treatment. The paper is designed to answer a narrower question: given a fixed $p_T^Z$ slice, does an ML-informed region in $(p_T^Z,m_{ZH})$ capture more of the benchmark SMEFT sensitivity than the corresponding vertical STXS region?

\section{Machine-learning-inspired region construction}
\label{sec:ml}

\subsection{Linear SVM in the \texorpdfstring{$(p_T^Z,m_{ZH})$}{(pTZ,mZH)} plane}
\label{subsec:svm}

Our most direct construction uses a linear support vector machine trained on the two variables $(p_T^Z,m_{ZH})$. We use the \texttt{LinearSVC} implementation in \textsc{scikit-learn} with $L_2$ regularization, tolerance $10^{-4}$, and a maximum of $10^4$ iterations. Event weights proportional to the physical cross sections are included during training.

The classifier returns a linear separator in the standardized feature space, which we rescale and express in physical units as
\begin{equation}
 f(p_T^Z)=a\,p_T^Z+b.
\label{eq:svm_line}
\end{equation}
One side of the line $m_{ZH} > f$ is defined as the EFT-enriched region and the other $m_{ZH} < f$ as the SM-like region. To maintain a clean comparison with the official STXS setup, we construct one such boundary inside each of the four standard $p_T^Z$ intervals of Eq.~\eqref{eq:stxsbins}. The resulting parameters are listed in Table~\ref{tab:svm_table}.

\begin{table}[t]
    \centering
    \begin{tabular}{c|cc}
        \hline
        STXS slice & Slope $a$ & Intercept $b$ [GeV] \\
        \hline
        $[75,150]$ & $-87 \pm 5$ & $9700 \pm 500$ \\
        $[150,250]$ & $-59 \pm 4$ & $11700 \pm 700$ \\
        $[250,400]$ & $-28.6 \pm 0.6$ & $9500 \pm 200$ \\
        $[400,\infty]$ & $-9.1 \pm 0.1$ & $6020 \pm 50$ \\
        \hline
    \end{tabular}
    \caption{Linear SVM boundaries obtained in each official STXS $p_T^Z$ slice. Uncertainties indicate the spread under the resampling procedure used in the analysis.}
    \label{tab:svm_table}
\end{table}

The physical content of the result is already visible here: the learned boundary is tilted, selecting events that are simultaneously hard in $p_T^Z$ and $m_{ZH}$, rather than merely large in one variable.

\subsection{Significance-driven refinement of the linear boundary}
\label{subsec:svm_scan}

The linear SVM is trained to maximize classification performance, not directly the significance metric of Eqs.~\eqref{eq:asimov_basic} and \eqref{eq:asimov_syst}. Since our aim is to optimize for the later, we therefore perform a local scan in slope and intercept around the SVM solution and select the boundary that maximizes the Asimov significance on the training set for a fixed choice of $\epsilon$.

\begin{figure}[t]
    \centering
    \includegraphics[width=0.75\textwidth]{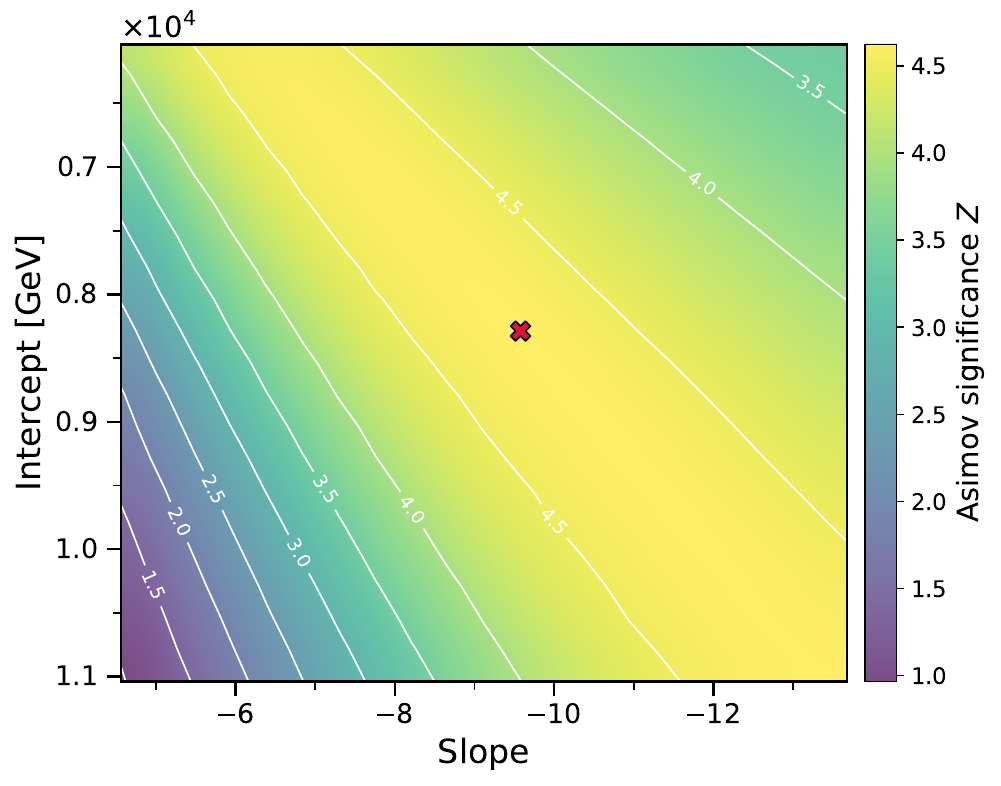}
    \caption{Asimov significance in the slope--intercept plane for the scan around the SVM solution in the highest-$p_T^Z$ slice, for a representative choice $\epsilon=0.5$.}
    \label{fig:svm_scan}
\end{figure}

Operationally, for each STXS slice and each value of $\epsilon$ we scan a family of lines around the SVM seed, compute the corresponding $(s,b)$ yields, evaluate $Z_A$, and retain the maximizing point. Figure~\ref{fig:svm_scan} shows the resulting significance landscape in the highest-$p_T^Z$ slice for $\epsilon=0.5$.

The optimized parameters are reported in Table~\ref{tab:svm_scan_table}. Beyond the numerical values reported in the table, the main point is the stability of the preferred orientation. Once significance is used as the objective instead of classification performance, the region remains aligned with the correlated high-energy tail.

\begin{table}[t]
    \centering
    \begin{tabular}{cc|cc}
        \hline
        STXS slice & $\sigma_B/B$ & Slope $a$ & Intercept $b$ [GeV] \\
        \hline
        $[75,150]$ & 0.3 & $-78 \pm 7$ & $11600 \pm 900$ \\
        $[75,150]$ & 0.5 & $-80 \pm 10$ & $12000 \pm 2000$ \\
        $[75,150]$ & 1.0 & $-59 \pm 10$ & $11000 \pm 1000$ \\
        $[150,250]$ & 0.3 & $-61 \pm 6$ & $15000 \pm 1000$ \\
        $[150,250]$ & 0.5 & $-56 \pm 6$ & $14000 \pm 1000$ \\
        $[150,250]$ & 1.0 & $-48 \pm 7$ & $13000 \pm 2000$ \\
        $[250,400]$ & 0.3 & $-33 \pm 3$ & $12000 \pm 1000$ \\
        $[250,400]$ & 0.5 & $-32 \pm 4$ & $12000 \pm 1000$ \\
        $[250,400]$ & 1.0 & $-24 \pm 2$ & $10200 \pm 800$ \\
        $[400,\infty]$ & 0.3 & $-10.0 \pm 0.9$ & $7300 \pm 600$ \\
        $[400,\infty]$ & 0.5 & $-9.5 \pm 0.3$ & $8200 \pm 200$ \\
        $[400,\infty]$ & 1.0 & $-8.0 \pm 0.8$ & $9000 \pm 700$ \\
        \hline
    \end{tabular}
    \caption{Significance-optimized SVM boundaries. The scan is performed separately for each STXS slice and each assumed fractional background uncertainty.}
    \label{tab:svm_scan_table}
\end{table}

\subsection{DNN-guided distillation into a linear region}
\label{subsec:dnn}

The second strategy asks whether a broader observable set materially changes the final low-dimensional region. We train a feed-forward DNN on the full feature set of Sec.~\ref{subsec:selection}. The network uses three hidden layers of sizes $(126, 84, 42)$ with ReLU activations, batch normalization after each hidden layer, Adam optimization with learning rate $10^{-4}$, weighted binary cross-entropy loss, and early stopping monitored on the validation accuracy.

The classifier output
\begin{equation}
s_{\rm DNN}(\mathbf{x})\in[0,1]
\end{equation}
is used only as an internal ranking variable. It is \emph{not} the final physics observable. Instead, for a threshold $t$ we select events with $s_{\rm DNN}(\mathbf{x})>t$, project that subset onto the $(p_T^Z,m_{ZH})$ plane, and fit a linear SVM to obtain a straight-line approximation of the high-score region. Scanning over $t$ therefore produces a family of linear boundaries,
\begin{equation}
 m_{ZH}=a_t\,p_T^Z+b_t,
\end{equation}
from which we choose the one that maximizes the Asimov significance for a given $\epsilon$.

\begin{figure}[t]
    \centering
    \begin{minipage}{0.45\linewidth}
        \centering
        \includegraphics[width=\linewidth]{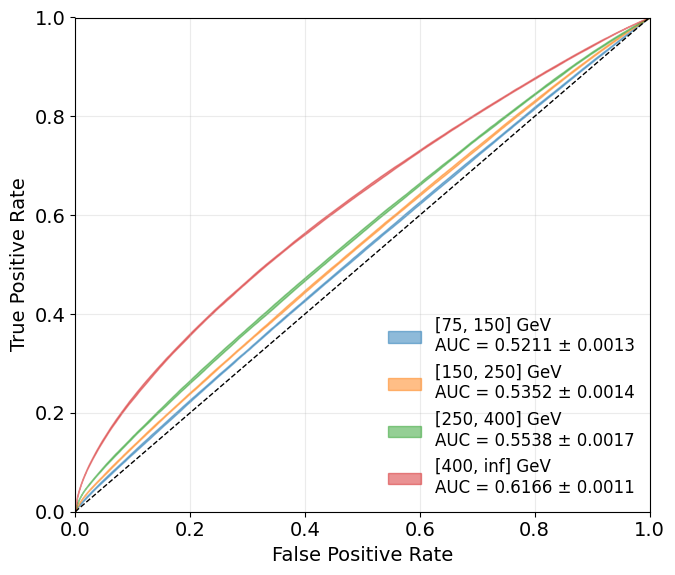}
    \end{minipage}
    \hfill
    \begin{minipage}{0.45\linewidth}
        \centering
        \includegraphics[width=\linewidth]{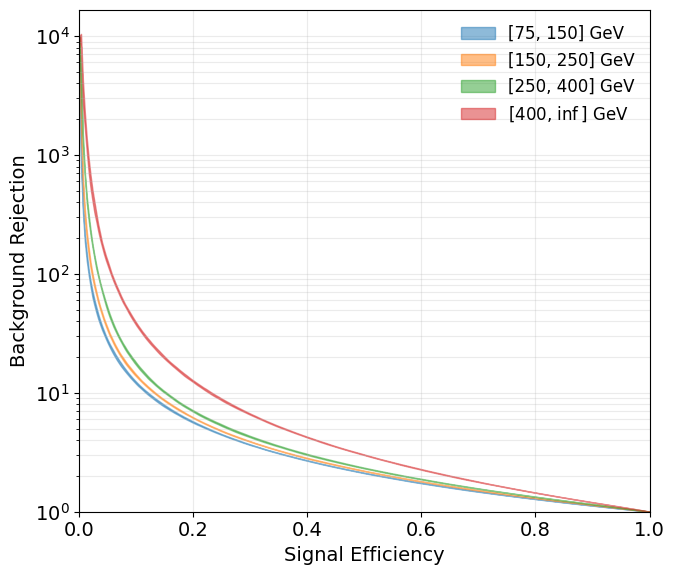}
    \end{minipage}
    \caption{Classifier performance inside the four official STXS $p_T^Z$ slices. Left: ROC curves for SM--benchmark discrimination with a shallow classifier trained separately in each slice. Right: background efficiency versus signal efficiency. The increasing separability toward large $p_T^Z$ motivates the focus on the boosted region.}
    \label{fig:roc_bin}
\end{figure}

\begin{figure}[t]
    \centering
    \includegraphics[width=0.75\textwidth]{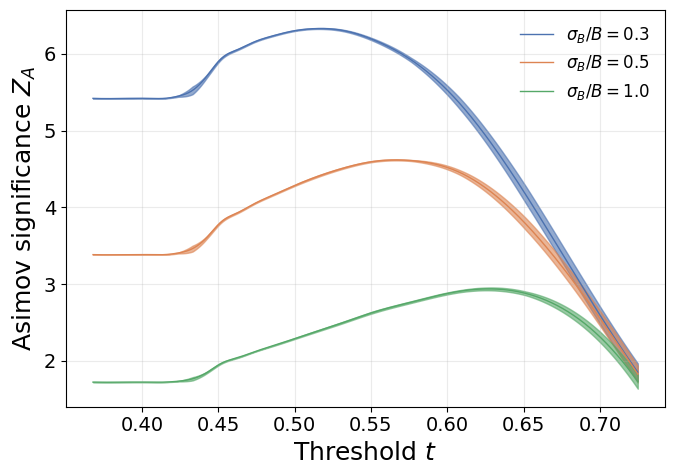}
    \caption{Asimov significance as a function of the DNN-score threshold used to define the projected high-score region in the highest-$p_T^Z$ slice.}
    \label{fig:dnnsvm_threshold}
\end{figure}

This procedure has two advantages. First, it lets the DNN exploit correlations beyond the two-dimensional plane. Second, it forces the final output back into a form that could plausibly be published as an STXS-like region definition. Figure~\ref{fig:roc_bin} shows the intrinsic classifier separation within each STXS slice, while Fig.~\ref{fig:dnnsvm_threshold} illustrates the threshold optimization for the highest-$p_T^Z$ bin.

The distilled boundary parameters are shown in Table~\ref{tab:dnnsvm_table}. In practice, they are steeper than the STXS vertical cuts and comparable to the SVM-based solutions, indicating that the DNN confirms rather than overturns the picture suggested by the two-dimensional training.

\begin{table}[t]
    \centering
    \begin{tabular}{cc|cc}
        \hline
        STXS slice & $\sigma_B/B$ & Slope $a$ & Intercept $b$ [GeV] \\
        \hline
        $[75,150]$ & 0.3 & $-80 \pm 40$ & $12000 \pm 5000$ \\
        $[75,150]$ & 0.5 & $-80 \pm 40$ & $12000 \pm 5000$ \\
        $[75,150]$ & 1.0 & $-70 \pm 40$ & $12000 \pm 5000$ \\
        $[150,250]$ & 0.3 & $-60 \pm 20$ & $14000 \pm 4000$ \\
        $[150,250]$ & 0.5 & $-60 \pm 20$ & $14000 \pm 4000$ \\
        $[150,250]$ & 1.0 & $-50 \pm 20$ & $13000 \pm 3000$ \\
        $[250,400]$ & 0.3 & $-35 \pm 4$ & $13000 \pm 1000$ \\
        $[250,400]$ & 0.5 & $-32 \pm 4$ & $12000 \pm 1000$ \\
        $[250,400]$ & 1.0 & $-28 \pm 4$ & $12000 \pm 1000$ \\
        $[400,\infty]$ & 0.3 & $-12 \pm 1$ & $8600 \pm 600$ \\
        $[400,\infty]$ & 0.5 & $-11.2 \pm 0.9$ & $9700 \pm 700$ \\
        $[400,\infty]$ & 1.0 & $-10.6 \pm 0.9$ & $10800 \pm 900$ \\
        \hline
    \end{tabular}
    \caption{Linear boundaries obtained by DNN-guided distillation into the $(p_T^Z,m_{ZH})$ plane.}
    \label{tab:dnnsvm_table}
\end{table}

\begin{figure}[t]
    \centering
    \includegraphics[width=0.75\textwidth]{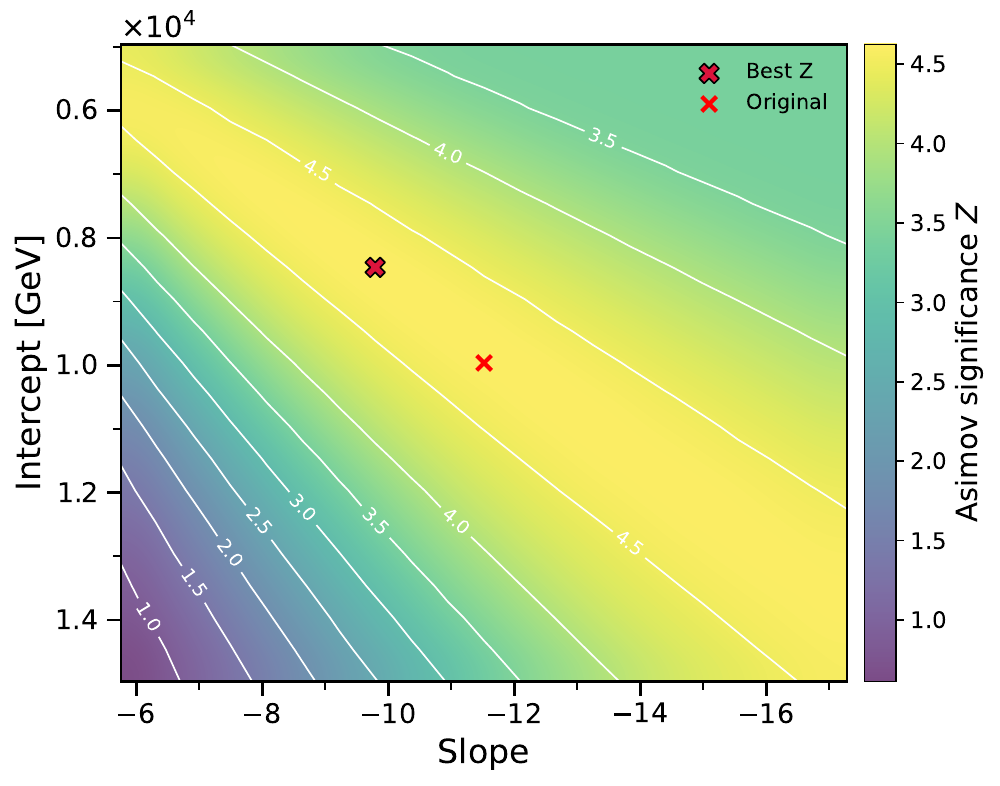}
    \caption{Significance landscape for the post-fit slope--intercept scan applied to the DNN-distilled boundary in the highest-$p_T^Z$ slice, for $\epsilon=0.5$.}
    \label{fig:dnnsvm_scan}
\end{figure}

For completeness we also apply the same post-fit slope--intercept scan used in the SVM case. The resulting parameters are listed in Table~\ref{tab:dnnsvmscan_table}, and a representative scan is shown in Fig.~\ref{fig:dnnsvm_scan}. The numerical changes are modest, which indicates that the threshold optimization already lands close to the significance optimum.

\begin{table}[t]
    \centering
    \begin{tabular}{cc|cc}
        \hline
        STXS slice & $\sigma_B/B$ & Slope $a$ & Intercept $b$ [GeV] \\
        \hline
        $[75,150]$ & 0.3 & $-80 \pm 30$ & $12000 \pm 5000$ \\
        $[75,150]$ & 0.5 & $-60 \pm 40$ & $11000 \pm 6000$ \\
        $[75,150]$ & 1.0 & $-60 \pm 30$ & $11000 \pm 5000$ \\
        $[150,250]$ & 0.3 & $-70 \pm 10$ & $16000 \pm 3000$ \\
        $[150,250]$ & 0.5 & $-55 \pm 6$ & $14000 \pm 2000$ \\
        $[150,250]$ & 1.0 & $-50 \pm 20$ & $13000 \pm 4000$ \\
        $[250,400]$ & 0.3 & $-33 \pm 5$ & $12000 \pm 2000$ \\
        $[250,400]$ & 0.5 & $-31 \pm 5$ & $12000 \pm 2000$ \\
        $[250,400]$ & 1.0 & $-23 \pm 5$ & $10000 \pm 2000$ \\
        $[400,\infty]$ & 0.3 & $-10.8 \pm 0.6$ & $7800 \pm 400$ \\
        $[400,\infty]$ & 0.5 & $-9.7 \pm 0.6$ & $8400 \pm 400$ \\
        $[400,\infty]$ & 1.0 & $-9 \pm 1$ & $10000 \pm 1000$ \\
        \hline
    \end{tabular}
    \caption{DNN-distilled boundaries after the additional significance-driven slope--intercept scan.}
    \label{tab:dnnsvmscan_table}
\end{table}

\section{Results}
\label{sec:results}

\subsection{Comparison of the learned boundaries}
\label{subsec:boundaries}

\begin{figure}[t]
    \centering
    \includegraphics[width=0.65\textwidth]{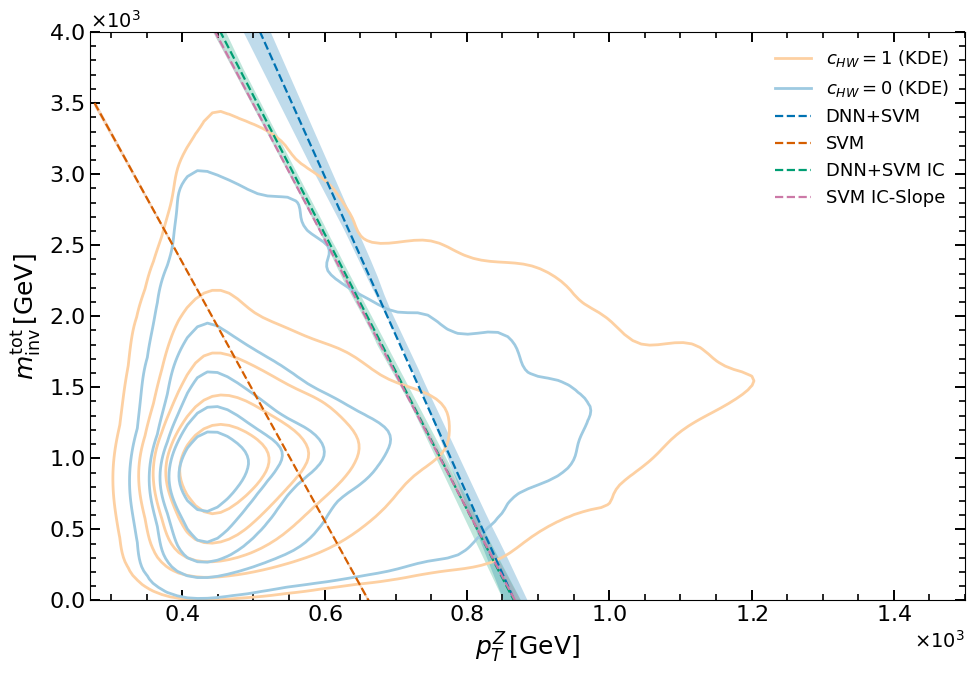}
    \caption{Comparison of the official STXS boundary with the ML-inspired linear boundaries in the highest-$p_T^Z$ slice, shown for the representative background-uncertainty choices used in the analysis. The ML-guided lines select the correlated high-$p_T^Z$, high-$m_{ZH}$ region more efficiently than the one-dimensional STXS slicing.}
    \label{fig:all_boundaries_fig}
\end{figure}

Figure~\ref{fig:all_boundaries_fig} compares, in the highest-$p_T^Z$ slice, the official STXS region boundary with the linear SVM and DNN-distilled alternatives. Several qualitative features are robust across methods. First, both ML-inspired constructions tilt toward the correlated hard tail, rather than following a vertical cut. Second, the optimized lines move toward more selective regions as the assumed fractional uncertainty increases, reflecting the fact that significance favors a purer region when the background is less well controlled. Third, the difference between the DNN-distilled boundary and the SVM-based one is noticeable but not dramatic. Once projected into $(p_T^Z,m_{ZH})$, the core geometric picture is already visible in the two-dimensional training.

This is the central geometric result of the paper. We explicitly set out to discover the most physically sensible correlated direction that can be encoded in a single straight line, instead of proposing an exotic nonlinear classifier.

\subsection{Single-region significance comparison}
\label{subsec:significance}

\begin{figure}
    \centering
    \includegraphics[width=0.75\linewidth]{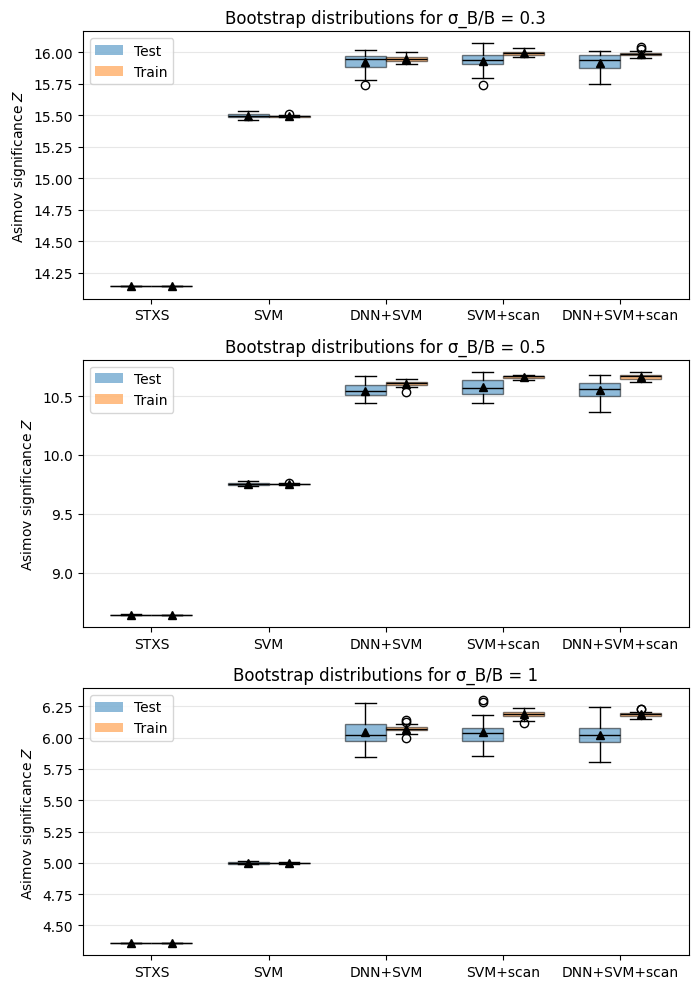}
    \caption{Distribution of the final Asimov significance yields for each method summed over all bin regions. The distributions are obtained from the bootstrap procedure on both sample set and models.}
    \label{fig:boxplots}
\end{figure}

Tables~\ref{tab:Z_03}--\ref{tab:Z_10} together with the boxplots in Figures~\ref{fig:boxplots} summarize the Asimov significance obtained with the official STXS region and with the ML-inspired alternatives inside each of the four $p_T^Z$ slices. The quoted uncertainties come from the bootstrap procedure used in the numerical study.
The three values of $\epsilon$ should be read as effective uncertainty levels motivated by the range currently encountered in $VH$ analyses, from relatively controlled regions to more uncertainty-dominated boosted configurations.

\begin{table}[h]
\centering
\small
\begin{tabular}{c|c|c|c|c}
\hline
Method & Bin 1 & Bin 2 & Bin 3 & Bin 4 \\
\hline
STXS & $2.244 \pm 0.001$ & $2.819 \pm 0.001$ & $3.656 \pm 0.001$ & $5.424 \pm 0.001$ \\
SVM & $2.393 \pm 0.010$ & $2.988 \pm 0.009$ & $3.911 \pm 0.013$ & $6.204 \pm 0.016$ \\
DNN+SVM & $2.529 \pm 0.045$ & $3.086 \pm 0.038$ & $3.983 \pm 0.024$ & $6.328 \pm 0.023$ \\
SVM+scan & $2.531 \pm 0.051$ & $3.084 \pm 0.047$ & $3.992 \pm 0.027$ & $6.324 \pm 0.023$ \\
DNN+SVM+scan & $2.515 \pm 0.050$ & $3.085 \pm 0.038$ & $3.991 \pm 0.024$ & $6.328 \pm 0.023$ \\
\hline
\end{tabular}
\caption{Asimov significance for $\sigma_B/B=0.3$. Each entry corresponds to a single-region comparison within a fixed STXS $p_T^Z$ slice.}
\label{tab:Z_03}
\end{table}

\begin{table}[h]
\centering
\small
\begin{tabular}{c|c|c|c|c}
\hline
Method & Bin 1 & Bin 2 & Bin 3 & Bin 4 \\
\hline
STXS & $1.347 \pm 0.001$ & $1.695 \pm 0.001$ & $2.216 \pm 0.001$ & $3.387 \pm 0.001$ \\
SVM & $1.438 \pm 0.006$ & $1.803 \pm 0.006$ & $2.401 \pm 0.008$ & $4.114 \pm 0.011$ \\
DNN+SVM & $1.514 \pm 0.051$ & $1.904 \pm 0.027$ & $2.519 \pm 0.019$ & $4.611 \pm 0.025$ \\
SVM+scan & $1.532 \pm 0.058$ & $1.907 \pm 0.032$ & $2.511 \pm 0.019$ & $4.631 \pm 0.026$ \\
DNN+SVM+scan & $1.494 \pm 0.077$ & $1.910 \pm 0.033$ & $2.518 \pm 0.018$ & $4.628 \pm 0.026$ \\
\hline
\end{tabular}
\caption{Asimov significance for $\sigma_B/B=0.5$.}
\label{tab:Z_05}
\end{table}

\begin{table}[h]
\centering
\small
\begin{tabular}{c|c|c|c|c}
\hline
Method & Bin 1 & Bin 2 & Bin 3 & Bin 4 \\
\hline
STXS & $0.674 \pm 0.0003$ & $0.849 \pm 0.0004$ & $1.113 \pm 0.0003$ & $1.724 \pm 0.0003$ \\
SVM & $0.720 \pm 0.003$ & $0.904 \pm 0.003$ & $1.212 \pm 0.004$ & $2.163 \pm 0.006$ \\
DNN+SVM & $0.806 \pm 0.082$ & $0.994 \pm 0.049$ & $1.330 \pm 0.017$ & $2.912 \pm 0.024$ \\
SVM+scan & $0.784 \pm 0.072$ & $0.973 \pm 0.050$ & $1.336 \pm 0.024$ & $2.952 \pm 0.027$ \\
DNN+SVM+scan & $0.772 \pm 0.091$ & $0.968 \pm 0.042$ & $1.328 \pm 0.023$ & $2.953 \pm 0.026$ \\
\hline
\end{tabular}
\caption{Asimov significance for $\sigma_B/B=1.0$.}
\label{tab:Z_10}
\end{table}

The linear SVM improves over the official STXS region in every slice and for every uncertainty choice shown. The gain is modest in the lowest-$p_T^Z$ region and grows steadily toward the boosted regime, exactly where the benchmark SMEFT deformation is expected to be most concentrated.

The DNN-distilled boundary usually improves further, especially in the third and fourth slices. For the representative choice $\epsilon=0.5$, the best ML-inspired boundary in the highest-$p_T^Z$ slice reaches $Z_A\simeq4.63$ compared with $3.39$ for the corresponding STXS region, an increase of about $37\%$. For $\epsilon=1.0$ the same comparison gives an increase of about $71\%$. The exact percentage is therefore uncertainty-model dependent, and it is more accurate to speak of a substantial gain in the boosted region than to quote a single universal number.

The post-fit slope-intercept scan produces only marginal changes relative to the threshold-optimized DNA-assisted result. This is useful methodologically because it shows that the gain does not rely on aggressive fine tuning of the final line parameters.

The larger bootstrap spread of the DNN-based methods in the lowest two slices suggests caution. The DNN still improves on the STXS baseline there, but the practical leverage is clearly concentrated in the high-energy bins. That is also the region where the proof-of-concept is most compelling physically.

\begin{figure}[t]
    \centering
    \includegraphics[width=\textwidth]{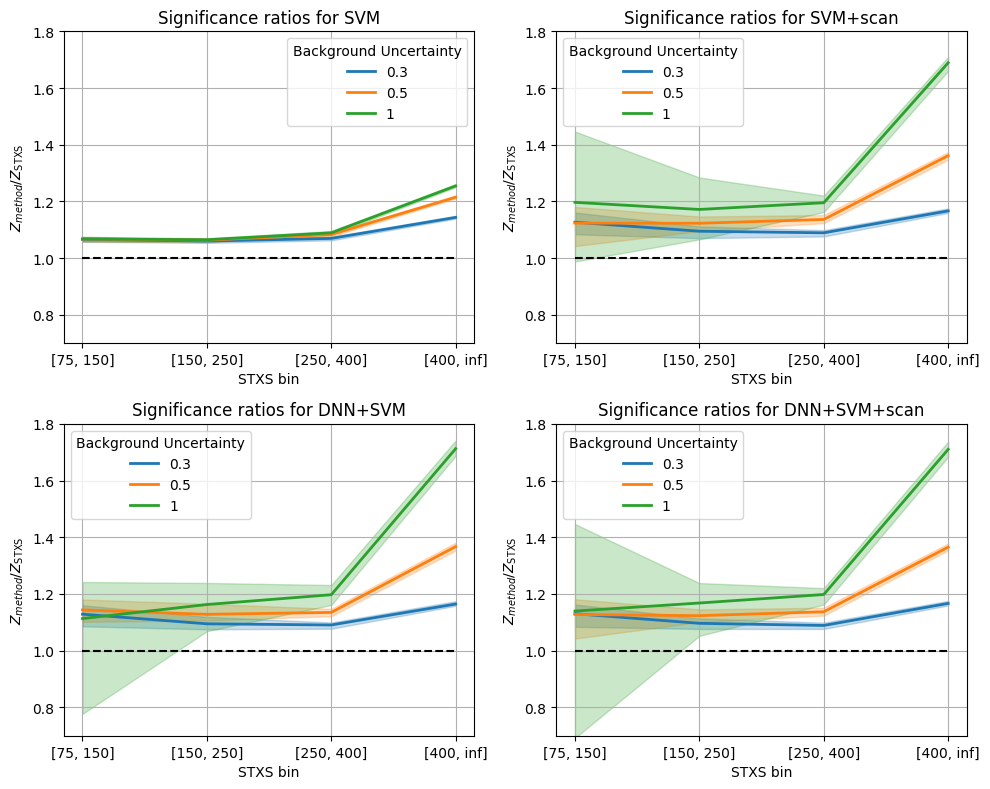}
    \caption{Ratio of the Asimov significance obtained with each ML-inspired region to that of the corresponding official STXS region, for the different $p_T^Z$ slices and uncertainty assumptions considered in the text.}
    \label{fig:ratios}
\end{figure}

The significance ratios relative to the STXS baseline are shown in Fig.~\ref{fig:ratios}. We recover the observations from before but relative to STXS: that significance yields increase in the high-energy regimes. However, the relative gain of significance in high background uncertainty scenarios is much greater, despite the overall yield being low. A familiar observation emerges when comparing the significance-optimized methods - SVM+scan, DNN+SVM and DNN+SVM+scan - to the simple SVM fit, for there is an increase in variance. This is markedly significant for the low bin and high background uncertainty case, in which some bootstrap experiments showed no improvement over STXS. We see overall that the advantage of the ML-inspired regions is systematic and becomes strongest precisely where the fixed one-dimensional slicing is least adapted to the signal morphology.

The results above support three claims and only these three claims.

First, the standard STXS segmentation is not optimally aligned with the benchmark SMEFT deformation studied here. Second, a simple linear cut in $(p_T^Z,m_{ZH})$ already recovers a nontrivial fraction of the multidimensional information that the vertical STXS slicing leaves unused. Third, adding more observables through a DNN helps, but the final low-dimensional region remains simple and interpretable.

What is \emph{not} established here is a full replacement of STXS, a channel-independent universal boundary, or a complete experimental sensitivity projection. The paper should be read as a targeted proof of concept showing that publishable region definitions can be improved with machine-learning guidance without becoming opaque.

\section{Conclusions}
\label{sec:conclusions}

We have presented a proof-of-concept extension of the STXS philosophy in which machine learning is used to \emph{design} simplified regions rather than to replace them with opaque classifier outputs. Using $pp\to ZH$ as a benchmark and focusing on a momentum-dependent bosonic SMEFT deformation, we showed that the relevant SM--SMEFT separation is organized along a correlated direction in $(p_T^Z,m_{ZH})$. The most attractive feature of the proposed construction is the fact that the final object is a low-parameter cut that can be documented, reproduced, and interpreted in exactly the way STXS regions are interpreted now, but with the phase-space direction backed up by a robust ML procedure.

A linear SVM trained directly on that plane already defines a region that systematically outperforms the corresponding official STXS region inside each $p_T^Z$ slice. A DNN trained on a broader observable set improves further, as much as a specialized significance-optimization of the direct SVM output. The fact that these two later approaches give similar results suggests that as long as the ML procedure takes significance optimization into account, the use of more kinetic observables in the DNN doesn't improve the significance yield much further. Seeing it from the other side, this result also shows that the gain does not rely on aggressive adjustment of the final line parameters, and so the DNN-assisted approach requires less fine-tuning.

For the representative background uncertainty choice $\sigma_B/B=0.5$, the gain in the highest-$p_T^Z$ slice is about $37\%$, and it is larger for more pessimistic uncertainty assumptions. The exact number is not universal, but the pattern is robust: the largest benefit appears precisely in the boosted regime where the benchmark SMEFT deformation is most concentrated. Likewise, we see that the regime where this happens is the higher $p_{TZ}$ bins, consistent with the enhancement of the hZZ vertex of high-energy regions in $ZH$ production.

For $ZH$ production, this separation works because the underlying physics is simple: the benchmark SMEFT deformation enhances a correlated boosted tail, and a straight line in $(p_T^Z,m_{ZH})$ is already a good approximation to the relevant geometry. The DNN confirms this picture rather than replacing it with a qualitatively different one. That is an encouraging sign for experimental usability. The next step is to test how far this philosophy generalizes across operators, across channels, and within more realistic experimental likelihoods. Even at its present level, however, this study makes a concrete point that if STXS is meant to be the bridge between precision measurements and EFT inference, then allowing machine learning to optimize the shape of that bridge is a natural next step. 




\acknowledgments

We thank members of the LHC EFT and Higgs working groups for valuable discussions.
This work is supported by the Spanish grants PID2023-148162NB-C21, CNS-2022-135688,
CEX2023-001292-S and MMT24-IFIC-01, which comes from the European Union's Recovery and Resilience Facility--Next Generation, in the framework of the General Invitation of the Spanish Government's public business entity Red.es to participate in talent attraction and retention programmes within Investment 4 of Component 19 of the Recovery, Transformation and Resilience Plan.
Computational resources were provided by the ARTEMISA cluster at the Universitat de Val\`encia.
The authors also acknowledge the developers of \textsc{MadGraph5\_aMC@NLO}, \textsc{SMEFTsim}, and open-source Python libraries including \textsc{scikit-learn}, \textsc{PyTorch}, and \textsc{NumPy}, used throughout this study.
\emph{Author contributions:} All authors contributed equally to this work.

\appendix

\section{Effective \texorpdfstring{$hZZ$}{hZZ} coupling from dimension-six operators}
\label{app:hZZcoupling}

In this appendix we collect the momentum-dependent $hZZ$ interaction induced by the bosonic SMEFT operators most relevant to associated Higgs production. In the Warsaw basis,
\begin{align}
\mathcal{O}_{HW} &= (H^\dagger H)\,W_{\mu\nu}^i W^{i\mu\nu}, &
\mathcal{O}_{HB} &= (H^\dagger H)\,B_{\mu\nu} B^{\mu\nu}, & \nonumber\\
\mathcal{O}_{HWB} &= (H^\dagger \tau^i H)\,W_{\mu\nu}^i B^{\mu\nu}.
\label{eq:O_HV}
\end{align}
After electroweak symmetry breaking, $H=(0,(v+h)/\sqrt{2})^T$, the terms linear in the Higgs field are
\begin{equation}
\mathcal{L} \supset \frac{v h}{\Lambda^2}\left[c_{HW}W^i_{\mu\nu}W^{i\mu\nu}+c_{HB}B_{\mu\nu}B^{\mu\nu}-\frac{1}{2}c_{HWB}W^3_{\mu\nu}B^{\mu\nu}\right].
\label{eq:L_vh}
\end{equation}
Using
\[
W^3_{\mu\nu}=c_w Z_{\mu\nu}+s_w A_{\mu\nu},
\qquad
B_{\mu\nu}=c_w A_{\mu\nu}-s_w Z_{\mu\nu},
\]
we obtain
\begin{equation}
\mathcal{L}\supset h\left[\kappa_{ZZ}Z_{\mu\nu}Z^{\mu\nu}+\kappa_{ZA}Z_{\mu\nu}A^{\mu\nu}+\kappa_{AA}A_{\mu\nu}A^{\mu\nu}\right],
\label{eq:L_hVV}
\end{equation}
with
\begin{align}
\kappa_{ZZ} &= \frac{v}{\Lambda^2}\Big(c_{HW}c_w^2+c_{HB}s_w^2+\tfrac{1}{2}c_{HWB}s_wc_w\Big),\\[3pt]
\kappa_{ZA} &= \frac{v}{\Lambda^2}\Big[2s_wc_w(c_{HW}-c_{HB})-\tfrac{1}{2}c_{HWB}(c_w^2-s_w^2)\Big],\\[3pt]
\kappa_{AA} &= \frac{v}{\Lambda^2}\Big(c_{HW}s_w^2+c_{HB}c_w^2-\tfrac{1}{2}c_{HWB}s_wc_w\Big).
\end{align}
The corresponding $hZZ$ vertex, with all momenta incoming, is
\begin{equation}
V^{\mu\nu}_{hZZ}(k_1,k_2)=i\,2\kappa_{ZZ}\Big[(k_1\!\cdot\!k_2)g^{\mu\nu}-k_1^{\nu}k_2^{\mu}\Big].
\label{eq:vertex_hZZ}
\end{equation}
The common Lorentz structure explains why this class of operators generically enhances correlated high-energy regions in $ZH$ production.

\section{DNN architecture and training details}
\label{app:dnn}

The DNN is used only as an intermediate representation from which a simple linear region can be distilled. It is therefore deliberately lightweight. The model is a fully connected multilayer perceptron with three hidden layers of sizes $(126,84,42)$ and ReLU activations. The output layer consists of a single sigmoid neuron producing a score $s_{\rm DNN}(x)\in[0,1]$.

The hidden-layer widths were chosen from the scaling pattern $(9m_{\rm hl},\,6m_{\rm hl},\,3m_{\rm hl})$, with the choice $m_{\rm hl}=14$ adopted in the final setup. Inputs are standardized to zero mean and unit variance. Training uses the Adam optimizer with learning rate $10^{-4}$ and weighted binary cross-entropy, where event weights encode the physical normalization of the SM and SMEFT samples. The network is trained for up to 200 epochs with batch size 1024, using early stopping on the validation loss with patience 3 and restoring the best weights.

After training, the DNN score is thresholded, the selected events are projected onto the $(p_T^Z,m_{ZH})$ plane, and a linear SVM is fitted to the selected subset. Scanning the threshold yields the family of DNN-guided linear boundaries discussed in Sec.~\ref{subsec:dnn}.

\section{Asimov significance}
\label{app:asimov}

For completeness, the Asimov-significance expressions used in the main text are reproduced here. For a single counting region with expected signal $s$ and background $b$, the profile-likelihood approximation without nuisance parameters gives
\begin{equation}
q_0=2\left[(s+b)\ln\left(1+\frac{s}{b}\right)-s\right],
\end{equation}
with $Z_A=\sqrt{q_0}$. Including a Gaussian background uncertainty $\sigma_b=\epsilon b$ leads to Eq.~\eqref{eq:asimov_syst} in the main text~\cite{Cowan:Asimov}.

\section{Principal-component analysis of the observable set}
\label{app:PCA}

To characterize the dominant directions of variation in the full observable set, we perform a principal-component analysis on the standardized features entering the DNN. The analysis is carried out separately in each official STXS $p_T^Z$ slice. The purpose is to check the expressivity of each variable in the high-dimensional dataset.

\begin{figure}[t]
    \centering
    \includegraphics[width=0.85\linewidth]{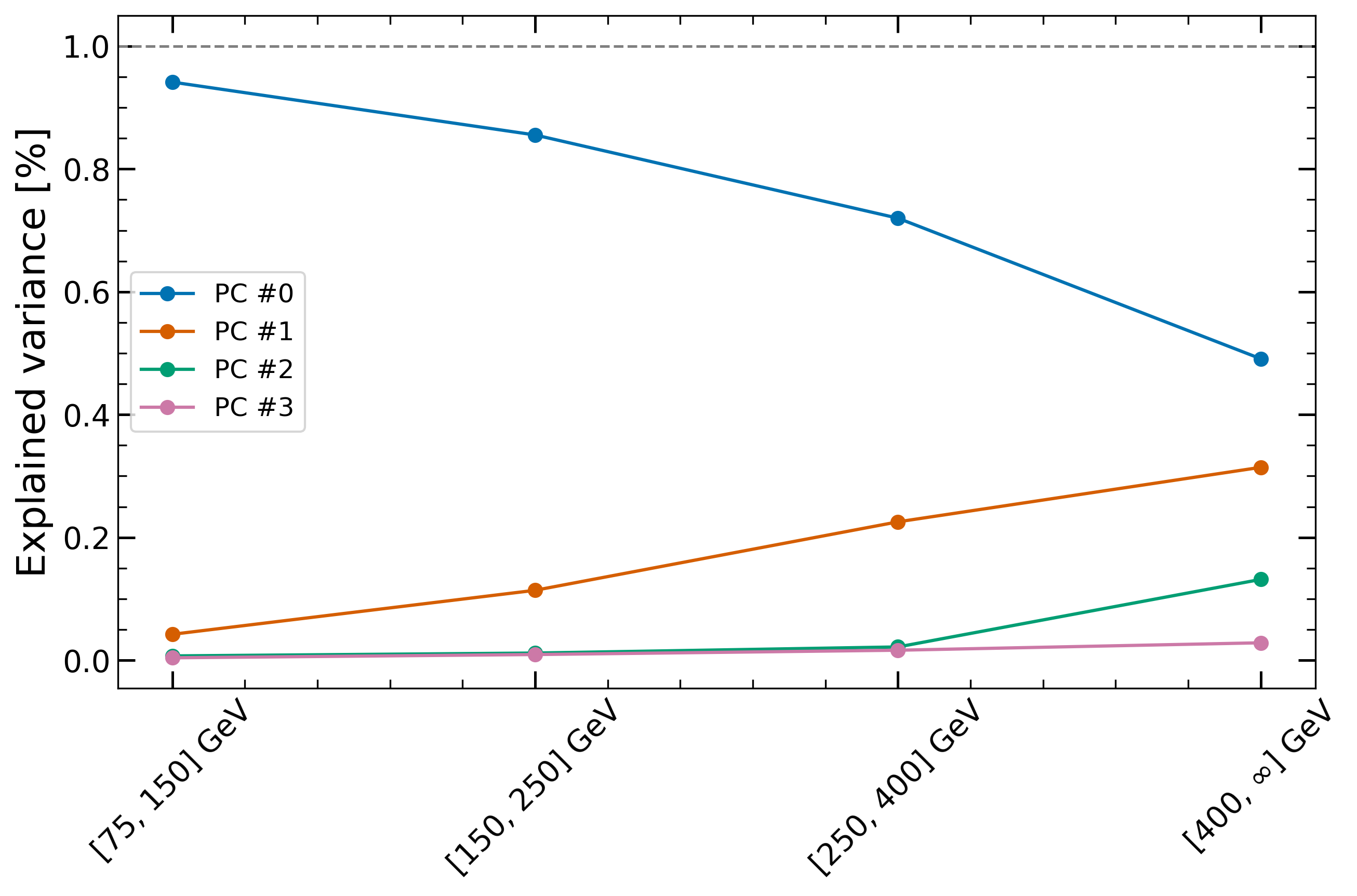}
    \caption{Fraction of explained variance carried by the first four principal components in each official STXS $p_T^Z$ slice. The leading component dominates throughout and becomes more pronounced in the boosted region.}
    \label{fig:pca_explained_variance}
\end{figure}

\begin{figure}[t]
    \centering
    \includegraphics[width=\linewidth]{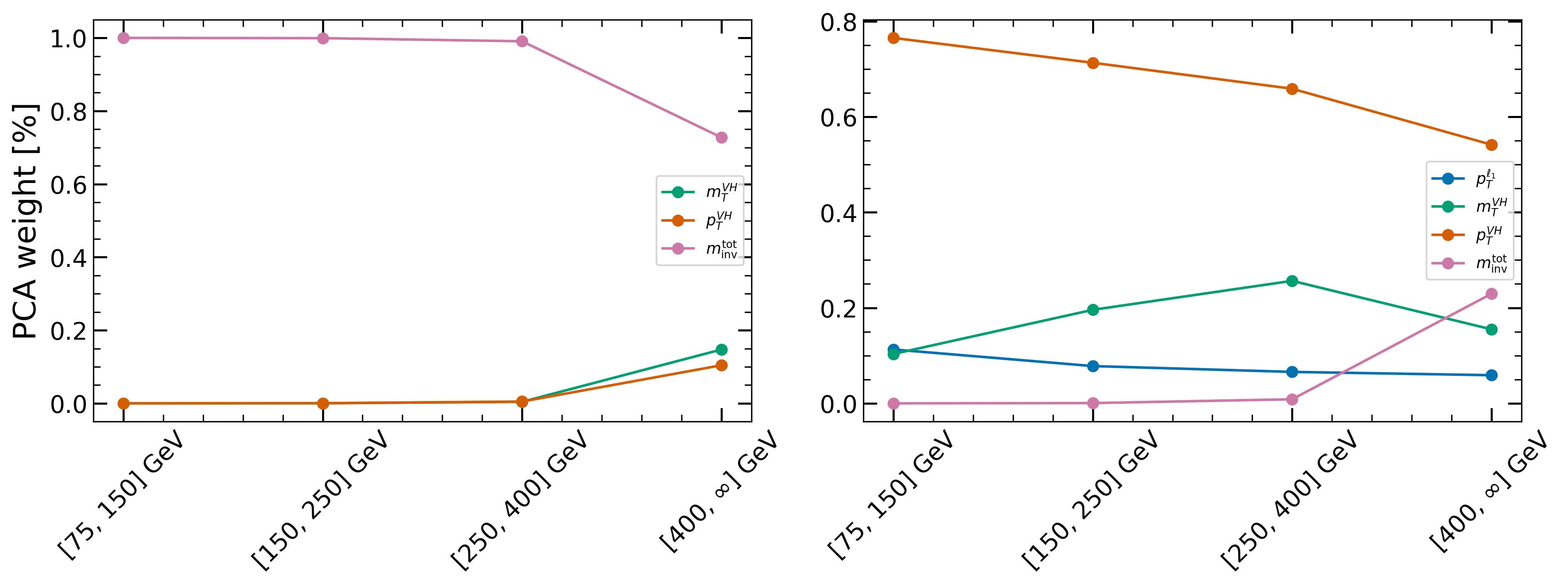}
    \caption{Principal-component weights for the two leading components, shown separately in each STXS slice. The first component is dominated by energy-like observables, supporting the use of $(p_T^Z,m_{ZH})$ as the plane in which to define the distilled linear boundary.}
    \label{fig:pca_weights}
\end{figure}

Figure~\ref{fig:pca_explained_variance} shows that the leading principal component dominates the variance in every slice, with increasing weight in the boosted regime. Figure~\ref{fig:pca_weights} shows that this leading component is driven primarily by invariant-mass- and transverse-momentum-related observables, while the next component carries more topological information. This is further motivation for a linear cut in $(p_T^Z,m_{ZH})$ which is expected to capture much of the useful structure.

\bibliographystyle{apsrev4-2}
\bibliography{mlstxs_refs}

@article{Ellis:2020unq,
    author = "Ellis, John and Madigan, Maeve and Mimasu, Ken and Sanz, Veronica and You, Tevong",
    title = "{Top, Higgs, Diboson and Electroweak Fit to the Standard Model Effective Field Theory}",
    eprint = "2012.02779",
    archivePrefix = "arXiv",
    primaryClass = "hep-ph",
    reportNumber = "KCL-PH-TH/2020-73, CERN-TH-2020-202",
    doi = "10.1007/JHEP04(2021)279",
    journal = "JHEP",
    volume = "04",
    pages = "279",
    year = "2021"
}

@article{Falkowski:2019hvp,
    author = "Falkowski, Adam and Straub, David",
    title = "{Flavourful SMEFT likelihood for Higgs and electroweak data}",
    eprint = "1911.07866",
    archivePrefix = "arXiv",
    primaryClass = "hep-ph",
    reportNumber = "LPT-Orsay-19-666",
    doi = "10.1007/JHEP04(2020)066",
    journal = "JHEP",
    volume = "04",
    pages = "066",
    year = "2020"
}

@article{Ellis:2012xd,
  author         = "Ellis, John and Sanz, Veronica and You, Tevong",
  title          = "{Associated Production Evidence for a Light Higgs Boson}",
  journal        = "Phys. Lett. B",
  volume         = "726",
  pages          = "244--250",
  year           = "2013",
  doi            = "10.1016/j.physletb.2013.08.048",
  eprint         = "1211.3068",
  archivePrefix  = "arXiv",
  primaryClass   = "hep-ph"
}

@article{Ellis:2013ywa,
  author         = "Ellis, John and Sanz, Veronica and You, Tevong",
  title          = "{The Effective Standard Model after LHC Run I}",
  journal        = "JHEP",
  volume         = "03",
  pages          = "157",
  year           = "2015",
  doi            = "10.1007/JHEP03(2015)157",
  eprint         = "1410.7703",
  archivePrefix  = "arXiv",
  primaryClass   = "hep-ph"
}

@article{Ellis:2014dva,
  author         = "Ellis, John and Sanz, Veronica and You, Tevong",
  title          = "{Complete Higgs Sector Constraints on Dimension-6 Operators}",
  journal        = "JHEP",
  volume         = "07",
  pages          = "036",
  year           = "2015",
  doi            = "10.1007/JHEP07(2015)036",
  eprint         = "1504.07200",
  archivePrefix  = "arXiv",
  primaryClass   = "hep-ph"
}

@article{Ellis:2014jta,
  author         = "Ellis, John and Sanz, Veronica and You, Tevong",
  title          = "{On the Interpretation of the LHC Higgs Results}",
  journal        = "JHEP",
  volume         = "06",
  pages          = "065",
  year           = "2015",
  doi            = "10.1007/JHEP06(2015)065",
  eprint         = "1404.3667",
  archivePrefix  = "arXiv",
  primaryClass   = "hep-ph"
}

@article{Alwall:2014hca,
    author = "Alwall, J. and Frederix, R. and Frixione, S. and Hirschi, V. and Maltoni, F. and Mattelaer, O. and Shao, H. -S. and Stelzer, T. and Torrielli, P. and Zaro, M.",
    title = "{The automated computation of tree-level and next-to-leading order differential cross sections, and their matching to parton shower simulations}",
    eprint = "1405.0301",
    archivePrefix = "arXiv",
    primaryClass = "hep-ph",
    reportNumber = "CERN-PH-TH-2014-064, CP3-14-18, LPN14-066, MCNET-14-09, ZU-TH-14-14",
    doi = "10.1007/JHEP07(2014)079",
    journal = "JHEP",
    volume = "07",
    pages = "079",
    year = "2014"
}

@article{Brivio:2017bnu,
    author = "Brivio, Ilaria and Trott, Michael",
    title = "{Scheming in the SMEFT... and a reparameterization invariance!}",
    eprint = "1701.06424",
    archivePrefix = "arXiv",
    primaryClass = "hep-ph",
    doi = "10.1007/JHEP07(2017)148",
    journal = "JHEP",
    volume = "07",
    pages = "148",
    year = "2017",
    note = "[Addendum: JHEP 05, 136 (2018)]"
}

@article{Ball:2017nwa,
    author = "Ball, Richard D. and others",
    collaboration = "NNPDF",
    title = "{Parton distributions from high-precision collider data}",
    eprint = "1706.00428",
    archivePrefix = "arXiv",
    primaryClass = "hep-ph",
    reportNumber = "EDINBURGH-2017-08, NIKHEF-2017-006, OUTP-17-04P, TIF-UNIMI-2017-3, CAVENDISH-HEP-17-06, CERN-TH-2017-077, Edinburgh 2017/08,
  Nikhef/2017-006, OUTP-17-04P,TIF-UNIMI-2017-3",
    doi = "10.1140/epjc/s10052-017-5199-5",
    journal = "Eur. Phys. J. C",
    volume = "77",
    number = "10",
    pages = "663",
    year = "2017"
}

@article{deFlorian:2016spz,
    author = "de Florian, D. and others",
    collaboration = "LHC Higgs Cross Section Working Group",
    title = "{Handbook of LHC Higgs Cross Sections: 4. Deciphering the Nature of the Higgs Sector}",
    eprint = "1610.07922",
    archivePrefix = "arXiv",
    primaryClass = "hep-ph",
    reportNumber = "CERN-2017-002-M, CERN-2017-002",
    doi = "10.23731/CYRM-2017-002",
    journal = "CERN Yellow Rep. Monogr.",
    volume = "2",
    pages = "1--869",
    year = "2017"
}

@article{ATLAS:STXS,
  author         = "{ATLAS Collaboration}",
  title          = "{Measurements of Higgs boson production cross sections and couplings using Simplified Template Cross Sections in the $H \to ZZ^* \to 4\ell$ decay channel at $\sqrt{s}=13$ TeV with the ATLAS detector}",
  journal        = "Phys. Lett. B",
  volume         = "786",
  year           = "2018",
  pages          = "59--86",
  doi            = "10.1016/j.physletb.2018.09.019",
  eprint         = "1808.09054",
  archivePrefix  = "arXiv"
}

@article{CMS:STXS,
  author         = "{CMS Collaboration}",
  title          = "{Measurements of Higgs boson production cross sections and couplings using Simplified Template Cross Sections in the $H\to \gamma\gamma$ decay channel at $\sqrt{s}=13$ TeV}",
  journal        = "Phys. Rev. D",
  volume         = "104",
  year           = "2021",
  pages          = "092013",
  doi            = "10.1103/PhysRevD.104.092013",
  eprint         = "2103.06956",
  archivePrefix  = "arXiv"
}

@article{Cowan:Asimov,
  author         = "Cowan, G. and Cranmer, K. and Gross, E. and Vitells, O.",
  title          = "{Asymptotic formulae for likelihood-based tests of new physics}",
  journal        = "Eur. Phys. J. C",
  volume         = "71",
  year           = "2011",
  pages          = "1554",
  doi            = "10.1140/epjc/s10052-011-1554-0",
  eprint         = "1007.1727",
  archivePrefix  = "arXiv"
}

@article{Azatov:2012bz,
  author         = "Azatov, Aleksandr and Contino, Roberto and Galloway, Jamison",
  title          = "{Model-Independent Bounds on a Light Higgs}",
  journal        = "JHEP",
  volume         = "04",
  pages          = "127",
  year           = "2012",
  doi            = "10.1007/JHEP04(2012)127",
  eprint         = "1202.3415",
  archivePrefix  = "arXiv",
  primaryClass   = "hep-ph"
}

@article{Azatov:2013hya,
  author         = "Azatov, Aleksandr and Contino, Roberto",
  title          = "{Boosting Higgs Signals with Energy Correlations}",
  journal        = "Phys. Rev. D",
  volume         = "88",
  number         = "7",
  pages          = "075019",
  year           = "2013",
  doi            = "10.1103/PhysRevD.88.075019",
  eprint         = "1308.2676",
  archivePrefix  = "arXiv",
  primaryClass   = "hep-ph"
}

@article{Biekotter:2018jzu,
  author         = "Biekötter, Thomas and Knochel, Alexander",
  title          = "{Energy-Enhanced Observables in Higgs and Diboson Production}",
  journal        = "JHEP",
  volume         = "07",
  pages          = "048",
  year           = "2018",
  doi            = "10.1007/JHEP07(2018)048",
  eprint         = "1805.00046",
  archivePrefix  = "arXiv",
  primaryClass   = "hep-ph"
}

@article{Giani:2023gfq,
    author = "Giani, Tommaso and Magni, Giacomo and Rojo, Juan",
    title = "{SMEFiT: a flexible toolbox for global interpretations of particle physics data with effective field theories}",
    eprint = "2302.06660",
    archivePrefix = "arXiv",
    primaryClass = "hep-ph",
    reportNumber = "Nikhef-2022-023",
    doi = "10.1140/epjc/s10052-023-11534-7",
    journal = "Eur. Phys. J. C",
    volume = "83",
    number = "5",
    pages = "393",
    year = "2023"
}

@techreport{Hays:2673969,
      author        = "Hays, Chris and Sanz Gonzalez, Veronica and Zemaityte,
                       Gabija",
      title         = "{Constraining EFT parameters using simplified template
                       cross sections}",
      institution   = "CERN",
      reportNumber  = "LHCHXSWG-2019-004",
      address       = "Geneva",
      year          = "2019",
      url           = "https://cds.cern.ch/record/2673969",
}

@article{Freitas:2019hbk,
    author = "Freitas, Felipe F. and Khosa, Charanjit K. and Sanz, Ver{\'o}nica",
    title = "{Exploring the standard model EFT in VH production with machine learning}",
    eprint = "1902.05803",
    archivePrefix = "arXiv",
    primaryClass = "hep-ph",
    doi = "10.1103/PhysRevD.100.035040",
    journal = "Phys. Rev. D",
    volume = "100",
    number = "3",
    pages = "035040",
    year = "2019"
}

@article{Brehmer:2018kdj,
  author         = "Brehmer, Johann and Cranmer, Kyle and Louppe, Gilles and Pavez, Juan",
  title          = "{A Guide to Constraining Effective Field Theories with Machine Learning}",
  journal        = "Phys. Rev. D",
  volume         = "98",
  number         = "5",
  pages          = "052004",
  year           = "2018",
  doi            = "10.1103/PhysRevD.98.052004",
  eprint         = "1805.00020",
  archivePrefix  = "arXiv",
  primaryClass   = "hep-ph"
}

@article{Cranmer:2015bka,
  author         = "Cranmer, Kyle and Pavez, Juan and Louppe, Gilles",
  title          = "{Approximating Likelihood Ratios with Calibrated Discriminative Classifiers}",
  journal        = "arXiv preprint",
  year           = "2015",
  eprint         = "1506.02169",
  archivePrefix  = "arXiv",
  primaryClass   = "stat.ML"
}

@article{Bishara:2017etm,
  author         = "Bishara, F. and Contino, R. and Rojo, J.",
  title          = "{Higgs pair production in vector-boson fusion at the LHC and beyond}",
  journal        = "Eur. Phys. J. C",
  volume         = "77",
  year           = "2017",
  pages          = "481",
  doi            = "10.1140/epjc/s10052-017-5037-7",
  eprint         = "1611.03860",
  archivePrefix  = "arXiv"
}

@article{Biekotter:2019kde,
  author         = "Biek{\"o}tter, A. and Knochel, A. and Kogler, R. and Plehn, T. and Voigt, A.",
  title          = "{Constraining effective field theories with machine learning}",
  journal        = "SciPost Phys.",
  volume         = "6",
  year           = "2019",
  pages          = "024",
  doi            = "10.21468/SciPostPhys.6.2.024",
  eprint         = "1812.07573",
  archivePrefix  = "arXiv"
}

@article{Brivio:2019myy,
  author         = "Brivio, I. and Corbett, T. and Trott, M.",
  title          = "{The SMEFTsim and HEPfit frameworks for global fits}",
  journal        = "JHEP",
  volume         = "10",
  year           = "2019",
  pages          = "056",
  doi            = "10.1007/JHEP10(2019)056",
  eprint         = "1906.11887",
  archivePrefix  = "arXiv"
}

@article{Brehmer:ML,
  author         = "Brehmer, J. and Cranmer, K. and Louppe, G. and Pavez, J.",
  title          = "{Constraining Effective Field Theories with Machine Learning}",
  journal        = "Phys. Rev. Lett.",
  volume         = "121",
  year           = "2018",
  pages          = "111801",
  doi            = "10.1103/PhysRevLett.121.111801",
  eprint         = "1805.00013",
  archivePrefix  = "arXiv"
}

@article{Kasieczka:NNinterpret,
  author         = "Kasieczka, G. and Nachman, B. and Shih, D.",
  title          = "{New Methods and Datasets for Deep Learning in High Energy Physics: A Review}",
  journal        = "Rept. Prog. Phys.",
  volume         = "84",
  year           = "2021",
  pages          = "124201",
  doi            = "10.1088/1361-6633/ac36b9",
  eprint         = "2101.05072",
  archivePrefix  = "arXiv"
}

@article{HEPfit,
  author         = "de Blas, J. and others",
  title          = "{Global SMEFT analysis using the HEPfit framework}",
  journal        = "JHEP",
  volume         = "12",
  year           = "2020",
  pages          = "135",
  doi            = "10.1007/JHEP12(2020)135",
  eprint         = "2007.04311",
  archivePrefix  = "arXiv"
}

@article{SFitter,
  author         = "Ellis, J. and You, T.",
  title          = "{Updated Global Analysis of Higgs Couplings}",
  journal        = "JHEP",
  volume         = "06",
  year           = "2013",
  pages          = "103",
  doi            = "10.1007/JHEP06(2013)103",
  eprint         = "1303.3879",
  archivePrefix  = "arXiv"
}

\end{document}